\documentclass{article}
\usepackage{arxiv}
\usepackage[affil-it]{authblk}
\usepackage[english]{babel}
\usepackage{graphicx} 
\usepackage{hyperref}
\usepackage[export]{adjustbox}
\usepackage{subcaption,lipsum}
\usepackage[labelfont=bf]{caption}
\usepackage{soul}
\usepackage{xcolor}
\usepackage{mathtools}

\DeclarePairedDelimiter\floor{\lfloor}{\rfloor}
\usepackage{amssymb}
\usepackage{amsmath}

\title{\textbf{TiMePReSt: Time and Memory Efficient Pipeline Parallel DNN Training with Removed Staleness}}

\author[1]{Ankita Dutta}
\author[2]{Nabendu Chaki}
\author[1,*]{Rajat K. De}

\affil[1]{Machine Intelligence Unit, Indian Statistical Institute, 203 Barrackpore Trunk Road, Kolkata 700108, India.}
\affil[2]{Department of Computer Science and Engineering, University of Calcutta, JD-2, Sector III, Salt Lake City, Kolkata 700098, India}
\affil[*]{Corresponding author: Rajat K. De, rajat@isical.ac.in}

\begin{document}
\maketitle
\begin{abstract}
DNN training is extremely time-consuming, necessitating efficient multi-accelerator parallelization, where a single iteration of training is split over the available accelerators. Current approaches often parallelize training by using intra-batch parallelization. Combining inter-batch pipeline parallelism with intra-batch parallelism is a common approach to further improve parallel training throughput. In this article, we develop a system, called TiMePReSt, that adds both of them, but in a different way which helps to better overlap computation and communication within a mini-batch, and limits the amount of communication. The traditional pipeline parallel training of DNNs maintains similar working principle as sequential or conventional training of DNNs. 
Thus, it suffers from high GPU memory footprint during training to maintain consistent version of weights in forward and backward passes of a mini-batch, similar to sequential training.Here, it has been shown experimentally that violating consistency of weight versions does not necessarily reduce prediction capability of a parallely trained DNN. TiMePReSt helps to overcome GPU memory overhead and achieve zero degree of staleness of weights by sacrificing sequential stability, but not effecting prediction capability. State-of-the-art pipeline parallel DNN training techniques often become costly in terms of training time. In order to address this issue, TiMePReSt introduces a variant of intra-batch parallelism that parallelizes the forward pass of each mini-batch by decomposing it into smaller micro-batches. Moreover, synchronization between backward and forward passes are performed in a novel way reduce training time in TiMePReSt. The chances of occurring multiple sequence problem and its relation with version difference have been observed in TiMePReSt. The version difference is observed to be varied with the number of micro-batches and worker machines. In this paper, a mathematical relationship between the number of micro-batches and worker machines has been formulated. Moreover, a mathematical expression of version difference has also been devised so that the version difference for different combination of these two can be computed mathematically without preparing diagrams for all the combinations.
\end{abstract}
\section{Introduction}
\label{Introduction}
Deep Neural Networks (DNNs) are highly effective and widely used in various domains like image classification, image recognition, language translation, language modeling, video captioning, speech recognition, and recommendation systems, among others. Due to its wide applicability, DNNs have grown massively in size to capture more complex scenarios, resulting in considerably complex computations during training. Nowadays, improved data generation technologies with low cost cause an exponential growth of data. Thus, more non-linear and large-sized datasets have become available, which are required by the increasingly complex DNNs to be trained on. As a result, the DNN training process becomes more expensive in terms of time and resources \cite{LIU2024317}. It is observed that with sufficient resources, DNNs can achieve dramatic improvements in its performance. A trained DNN is often used for inference which is not such costly task like training. The massive surge in resource requirements for DNN training has led to new devices such as Google’s TPU, NVidia’s Tesla, or Xilinx Alveo in addition to custom accelerators \cite{unnikrishnan2021layerpipe}. Due to lack of cost-effectiveness of those high-end servers, parallelisation of the training process over multiple commodity hardwares has become relevant, despite the fact that most deep learning engineers and researchers are used to write non-distributed code. It is reasonably difficult for them to adapt to parallel and distributed programming. 
\\\\Data parallelism \cite{LI2021206} is the simplest among all the parallelism approaches in terms of implementation, since each machine in a cluster has the entire DNN to be trained in memory. The training dataset is divided and distributed over the cluster so that each machine performs forward and backward passes of the DNN on the allocated chunk of data parallely and independently. Finally, the weights are updated based on the collective effect of all independent runs of forward-backward sequences. Model parallelism is complicated compared to data parallelism in terms of implementation. Model parallelism \cite{LI2021206} is based on the concept of distributing a DNN to be trained rather than the dataset over the cluster. A DNN can be divided either node(neuron)-wise or layer-wise. These two variants are conceptually known as tensor parallelism and pipeline parallelism respectively. In contrast to data parallelism, model parallelism does not distribute training data over the cluster. Depending on the problem scope, it can be decided which machines need to store the data locally. Model parallelism is capable of handling large networks which data parallelism cannot, whereas both of them can handle large data size in their own way.
\\\\In this paper, we have designed a model, called Time and Memory Efficient Pipeline Parallel DNN Training with Removed Staleness (TiMePReSt) to train DNNs on a pipeline parallel system, distributing the entire network layer-wise across multiple accelerators (either in the same machine or multiple machines). TiMePReSt is advantageous over the existing systems in terms of $1)$ the extent of staleness of the trainable parameters, $2)$ time efficiency and $3)$ memory efficiency.
TiMePReSt is able to execute both forward and backward passes of a single mini-batch with the most recent version of weights, in contrast to the state-of-the-art methods. More specifically, it compromises consistency of weight versions in forward and backward passes of a mini-batch in order to utilize the weight update, which was yet to occur while the forward pass starts. Additionally, it does not require to store an older version of weights until a mini-batch completes both the forward and backward passes. The memory can be freed up once an updated version is available and no mini-batch is still using the older version. Thus, removal of horizontal weight stashing \cite{narayanan2019pipedream} makes the model more memory-efficient. As a result of the version inconsistency, the number of iterations required for achieving a particular accuracy increases. In other words, convergence is achieved after more number of epochs, which acts as a bottleneck for time-consuming training process of DNNs. TiMePReSt is able to address the time-inefficiency problem since it completes a training epoch in significantly less time compared to PipeDream. Thus, TiMePReSt is able to afford extra epochs in a given amount of time, compared to PipeDream. It has been shown experimentally that TiMePReSt is able to achieve a particular accuracy much faster than PipeDream. The reason behind the time-efficiency of TiMePReSt over PipeDream is reduction of the occurrence of the most time-consuming tasks in the entire training process. Backward propagation of the prediction errors and updating weights are the most computaionally expensive and time-consuming tasks. TiMePReSt divides each mini-batch in a number of micro-batches, and performs forward passes for each of them. However, both of the above tasks are performed only once in each mini-batch. It has been shown experimentally that the novel kind of above mentioned intra-batch parallelism technique is able to capture the collective effect of all the forward passes for the micro-batches very well in a single backward pass and weight update, without loosing the prediction capability.
\\\\In TiMePReSt, the likelihood of occurring multiple sequence problem and its relationship with version difference have been observed. The numbers of worker machines and micro-batches are observed to have a varying effect on the version difference. A mathematical relationship between the numbers of worker machines and micro-batches has been developed in this paper. Additionally, a mathematical expression for the version difference has been developed in terms of the numbers of workers and micro-batches so that, without creating diagrams for every possible combination, the version difference for each combination of these two can be calculated mathematically.
\section{Background}
In this section, we briefly describe different types of parallelism adopted so far for training deep neural networks. 
\subsection{Data Parallelism}
\begin{figure}[h]
  \begin{subfigure}{.58\textwidth}
  \centering
    \includegraphics[width=.9\linewidth]{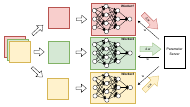}
    \caption{Parameter Server}
    \label{fig:Data Parallelism Parameter Server}
  \end{subfigure}
  \begin{subfigure}{.4\textwidth}
  \centering
    \includegraphics[width=.9\linewidth]{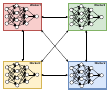}
    \caption{AllReduce}
    \label{fig:Data Parallelism all_reduce}
  \end{subfigure}
  \caption{\textbf{Different Model Synchronization Strategies in Data Parallelism}}
  \label{Different Model Synchronization Strategies in Data Parallelism}
\end{figure}

\begin{figure}[h]
  \begin{subfigure}{.5\textwidth}
  \centering
    \includegraphics[width=\linewidth]{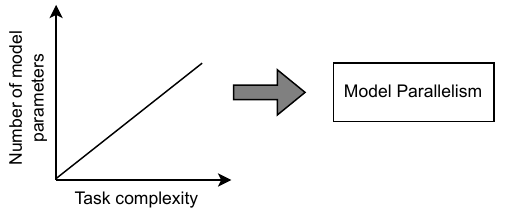}
    \caption{Model Parallelism}
    \label{Model Parallelism}
  \end{subfigure}
  \begin{subfigure}{.5\textwidth}
  \centering
    \includegraphics[width=\linewidth]{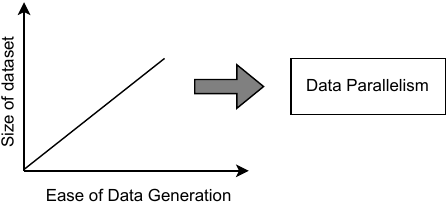}
    \caption{Data Parallelism}
    \label{Data Parallelism}
  \end{subfigure}
  \caption{\textbf{Scenarios where different parallelism techniques needed. Model Parallelism is preferred when to train a complex DNN with a number of parameters and complex architecture. Data parallelism is used to handle tremendous size of data.}}
  \label{Scenarios where different parallelism techniques needed}
\end{figure}

There exist several attempts to distribute the load of large input datasets into multiple processors or computers to compute gradient updates locally and then aggregating them to compute final update. The entire network is replicated to all workers. Each worker performs forward and backward passes on the same shard of data allocated to it, independent to other workers. Before the weight update phase, the gradients computed by the partitions have to be synchronized to obtain the final gradient for the entire training dataset across workers. The sync is performed based on the principle that - the gradient of a sum is same as the sum of the gradients. Furthermore, all DNN parameters are accessible to all workers, due to replication of the network \cite{shallue2019measuring}\cite{ben2019demystifying}. This phenomenon is known as data parallelism or pattern parallelism \cite{ben2019demystifying}. 
One of the earliest attempts of DNN training in data parallel architectures was due to Raina et al. \cite{raina2009large}. PyTorch \cite{paszke2019pytorch} provides a DistributedDataParallel (DDP) \cite{li2020pytorch} module to parallelize DNN training across multiple processors and machines.
\\\\In data parallelism, after each iteration, the model parameters have different values for different replicas. This is a common problem in data parallelism, which is termed as Model Synchronisation and is usually handled by two strategies, called Parameter Server Architecture \cite{dean2012large}\cite{li2013parameter} and All-Reduce Architecture, introduced by Baidu Research\footnote{\url{https://github.com/baidu-research/baidu-allreduce}}, illustrated in Figures \ref{fig:Data Parallelism Parameter Server} and \ref{fig:Data Parallelism all_reduce} respectively. In Parameter Server Architecture, there are one (or more) nodes other than the worker nodes, dedicated to store the gradients from all the workers and update the model parameters globally. GeePS \cite{cui2016geeps} is a parameter server system for scalable deep learning on distributed GPUs. In contrast, the All-Reduce Architecture does not have any centralised storage for gradients. The worker nodes individually compute the gradients locally, and synchronize the gradients from others. Data parallelism suffers from memory redundancy caused by each device having a duplicate copy of the model parameters, optimizer states, and gradients \cite{li2023colossal}.
\\\\Number of workers increase with the amount of input data since more partitions are required. Partitioning the data is needful to accelerate the training for small networks only. Since large networks increase number of model weights, thus, large amounts of data needs to be communicated between workers all at once. The amount of data communicated is proportional to the number of model weights and the number of workers participating in training \cite{narayanan2019pipedream}. Thus, communication overhead increases. Data parallelism is effective when the network size is small enough to have a replica of the entire network in each machine in a cluster. In cases where any single worker in a cluster is not capable of having a replica of the entire network, it is needed to divide the network into multiple machines. Model partitioning or model parallelism allows the network to be distributed among several devices, which is another method to fulfill the increasing computational demands. Figures \ref{Model Parallelism} and \ref{Data Parallelism} illustrate when it is needed to use data and model parallelisms.

\subsection{Model Parallelism}
As we have discussed earlier, both the computational load and memory footprint for training and inference rise proportionately with the complexity of the DNNs \cite{chilimbi2014project}. The increased number of operations required to complete forward and backward passes during training is the cause of the rise in computational needs. There are more mathematical operations in more complex models since they typically contain more layers, neurons, or sophisticated architectures. Similarly, larger models require more parameters, and each parameter needs to be stored in memory, which results in an increase in the memory footprint. Memory is also used by intermediate values produced during computation, and the number of these values increases with model complexity. Model parallelism \cite{brakel2024model} may be able to satisfy the enormous computational demands of DNNs. Tensor parallelism and layer-wise pipeline parallelism are two concepts within model parallelism.

\subsubsection*{\textit{Tensor Parallelism}}
\begin{figure}[h]
    \centering
    \includegraphics[width=0.5\textwidth, center]{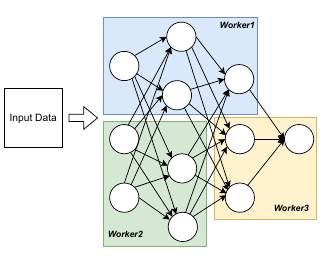}
    \caption{\textbf{An example Tensor Parallelism with three worker nodes. A DNN with four layers are distributed across the nodes based on intra-layer partitions.}}
    \label{fig:Tensor Parallelism}
\end{figure}

Tensor parallelism, also known as distributed tensor computation \cite{shoeybi2019megatron}, parallelizes computation within an computation-intensive operation such as matrix multiplication. In the context of DNNs, it is commonly referred to as intra-layer parallelism \cite{yi2022optimizing}, where each layer is divided and distributed into multiple devices. Thus, the entire computation to be performed in each layer is distributed accordingly. Figure \ref{fig:Tensor Parallelism} shows that a DNN is distributed across three different accelerators, and the partioning is done within layers. The nodes of each layer are distributed across machines. Megatron-LM \cite{shoeybi2019megatron} is an example of tensor-parallelism, which has been experimented to partition large transformer models over multiple GPUs. MegaScale \cite{jiang2024megascale} is another one, which has allowed training a large transformer model on 12,288 GPUs. Tensor parallel training technique splits a model “horizontally” within a layer. It is also possible to “vertically” split a model by layer in pipeline parallelism.
\subsubsection*{\textit{Pipeline Parallelism}}
\begin{figure}[h]
    \centering
    \includegraphics[width=0.5\textwidth, center]{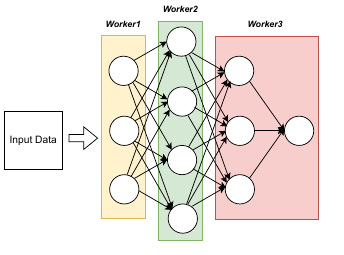}
    \caption{\textbf{An example Pipeline Parallelism with three worker nodes. A DNN with four layers are distributed layer-wise across the nodes. Each mini-batch of input data is passed through all the consecutive stages.}}
    \label{fig:Pipeline Parallelism}
\end{figure}

Pipeline parallelism is also known as layer-wise parallelism \cite{AKINTOYE2023432} or inter-layer parallelism. As shown in Figure \ref{fig:Pipeline Parallelism}, pipeline parallelism allows to decompose the network layer-wise and allocate the partitions to different GPUs. Each partition contains one or more consecutive layers. The computations are also distributed accordingly and the computation result of a partition is communicated to next device in pipeline. 
In order to build an efficient pipeline-based distributed DNN training framework, the key questions that need to be answered are: 1) Which version of weights should be used for the next mini-batches while earlier mini-batches are still executing in different stages and weight parameters are being updated by them? 
 What should be the permissible degree of staleness \cite{chen2016revisiting} of weight parameters in such scenario? 2) How can one overcome the bottleneck like time-inefficiency, a common drawback of any pipeline parallelism-based DNN training technique? 
 3) How can one overcome GPU memory overhead during training and how to make efficient usage of GPU memory?
\\\\Data parallelism does not perform versioning of weights since weight parameters are updated only once at the end of an epoch. However, weight versioning within an epoch is an important criterion for convergence of DNNs while training in distributed framework based on model parallelism, more specifically pipeline parallelism. Pipeline parallelism involves several weight updations within a single epoch. The weights are updated commonly after each mini-batch ends. However, the frequency of updation and way to update are very important research problems. In pipeline parallelism, mini-batches suffer from stale weights due to multiple updates within epoch. The degree of staleness also matters for convergence and high performance. Pipeline parallelism suffers from communication overhead for larger mini-batches, which makes it inefficient specially while running in a cluster consisting of multiple single GPU machines, since it involves communication between two computers followed by the GPUs. Due to lack of multi-GPU machines, we have opted for such a cluster. However, we have addressed the communication overhead issue without compromising with mini-batch size. 
\section{Related Works}
\label{Related Works}
\begin{figure}[ht]
    \centering
    \includegraphics[width=1\textwidth, center]{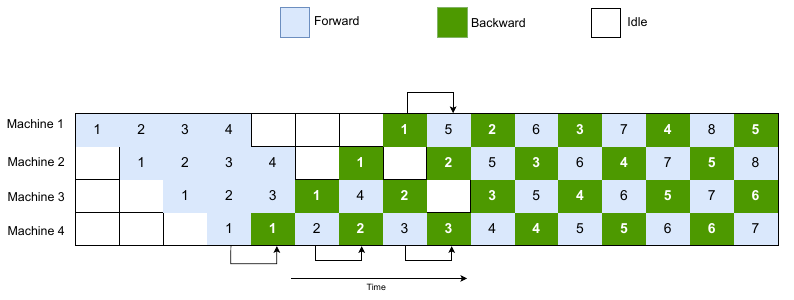}
    \caption{\textbf{An example of PipeDream parallel training with four workers. Each mini-batch maintains the same weight versions on both forward and backward passes.}}
    \label{fig:PipeDream}
\end{figure}

PipeDream \cite{narayanan2019pipedream} and GPipe \cite{huang2019gpipe}\cite{ZHANG2023107} are two contemporary as well as widely cited pipeline parallelism frameworks for large scale distributed deep learning. PipeDream combines inter-batch parallelism with intra-batch parallelism to increase pipelining throughput. It has introduced horizontal and vertical weight stashing to maintain same weight versions between forward and backward passes of a mini-batch, and those across the stages of a pass respectively. PipeDream is an asynchronous technique, which has devised a bi-directional training of DNNs, where a backward pass starts immediately after a forward pass finishes, through the same set of workers in reverse order. In order to prevent circular waiting between forward and backward passes, 1F1B (1 Forward 1 Backward) scheduling algorithm has been introduced in PipeDream. PipeDream injects multiple mini-batches at a time in the system for parallel training purpose. It follows the weight update rule given by
\begin{equation}\label{eqn:pipedream}
    \textbf{W}(t+1)=\textbf{W}(t) - \eta \cdot \nabla f\Bigl(\mathbf{W}_1(t-n+1), \mathbf{W}_2(t-n+1), \ldots, \mathbf{W}_n(t-n+1)\Bigr)
\end{equation}
Here, we denote $\mathbf{W}_l(t)$ as weights of the layers in $l^{th} (l = 1,2,...,n)$ stage after $t$ mini-batches, $\eta$ as the learning rate, $\nabla f\Bigl(\mathbf{W}_1(t), \mathbf{W}_2(t), \ldots, \mathbf{W}_n(t)\Bigr)$ is the gradient computed over all samples in the mini-batch, $f$ is the loss function and $\mathbf{W}(t)$ is the weights of the entire network (rather than a stage) after $t$ mini-batches.  Equation \ref{eqn:pipedream} shows that it computes gradients on stashed weights $\mathbf{W}_l(t-n+1)$. Thus, it suffers from staleness of weights due to horizontal and vertical weight stashing \cite{narayanan2019pipedream}, although the computed gradients are applied on different weight versions from those used to compute the gradients. Figure \ref{fig:PipeDream} shows that mini-batch 3 continues backward pass on the same version of weights, which it has used in forward pass, although an updated version of weights (updated by mini-batch 2) is already available before the backward pass starts. In addition, PipeDream is able to automatically partition the DNN optimally across the workers.
\\\\In contrast, GPipe \cite{huang2019gpipe} is synchronous, which implements uni-directional training of DNNs. In other words, backward passes can start only when forward passes of all the mini-batches end. Thus, no external scheduling is needed for the forward and backward passes. Moreover, GPipe is only applicable for the networks that can be expressed as a sequence of layers. Both GPipe and PipeDream allow different workers to process different input batches in parallel by providing numerous batches sequentially through a series of workers. Each worker controls one partition of the DNN. GPipe employs periodic pipeline flushes in order to maintain single weight version at a time, whereas PipeDream needs to have multiple weight versions at a time; thus, no periodic flushes occur. This property makes GPipe more memory efficient, however less time efficient than PipeDream since pipeline flushes decrease both memory footprint as well as pipeline throughput. 
\\\\Chen et al. \cite{chen2018efficient} have developed SpecTrain, which addresses 
staleness of weights, due to which prediction accuracy of the parallely trained DNN is compromised. SpecTrain adopts a future weight prediction method in early pipeline stages, which have been used by the mini-batches rather than the stale weights. The prediction is done based on the smoothed gradients used in momentum-based optimizers \cite{kingma2014adam} and the staleness computed. 
\\\\Guan et al. \cite{guan2019xpipe} have been introduced XPipe, which follows an asynchronous model parallel training approach, where each mini-batch is divided into a few micro-batches. The forward and backward passes of the micro-batches are performed although weights are updated only after the backward passes of all the micro-batches corresponding to a mini-batch complete. Since it allows cross-training of multiple mini-batches, it suffers from the weight inconsistency and staleness issues. In order to address this issue, the authors have formulated a metric to mathematically calculate the staleness, and XPipe performs weight update based on it.
\\\\Narayanan et al. \cite{narayanan2021memory} have developed two memory-efficient variants of PipeDream, named as PipeDream-2BW and PipeDream-Flush, which achieve high throughput also. In PipeDream-2BW \cite{narayanan2021memory}, the authors have introduced double-buffered weight updates (2BW), which reduces memory-footprint by reducing number of active weight versions and increases pipeline throughput by avoiding pipeline flushes. PipeDream-2BW maintains only two weight versions at a time for already in-flight inputs, and for newly admitted inputs. The authors have called the earlier one as shadow version. PipeDream-2BW performs single weight update for a series of inputs rather than individual update for all. However, it computes gradient for each of them and stores it into a “coalesced” gradient for further use. As a result, a constant weight delay of unity has been introduced implicitly. PipeDream-Flush \cite{narayanan2021memory} is another variant of PipeDream, which has less memory footprint than PipeDream-2BW in cost of pipeline throughput since it employs periodic pipeline flush in PipeDream.
\\\\Li et al. \cite{li2021chimera} have introduced Chimera that runs fully-packed bidirectional pipelines through the same set of accelerators for large-scale DNN training. Chimera combines two pipelines in opposite directions, together called - up and down pipelines. In both the cases, the stages are mapped to the workers in completely opposite order. For example, if there are three stages $s_0$, $s_1$, and $s_2$, then the stages are mapped in the order $s_0 - s_1 - s_2$ to the workers $w_2 - w_1 - w_0$ and $w_0 - w_1 - w_2$ in up and down pipelines respectively. In each direction, it allows $N/2$ micro-batches to be pushed before a pipeline flush, where $N$ denotes total number of training samples.
\\\\Boral et al. \cite{boral2023anomaly} have employed pipeline parallel DNN training technique for anomaly detection \cite{nassif2021machine} in streaming environment. A streaming environment is made up of continuously updated, high-dimensional data produced by IoTs, sensors, and other devices. In contrast to static environments, streaming environments enable machine learning algorithms to learn concepts in real time. The existence of these anomalies in the streaming environment can bottleneck real-time learning. Thus, anomaly detection is a crucial task.
\section{Methodology}
\label{Methodology}
In this work, we overcome two major challenges of the existing pipeline parallelism techniques. One is staleness of weights, and the other is huge training time. The reason behind staleness of weights is that during forward pass of a mini-batch, another mini-batch can make update to weights, which makes the earlier version of weights stale. In order to overcome stale or delayed weight problem in pipeline parallelism, the proposed TiMePReSt incorporates relaxation by allowing different weight versions in forward and backward passes corresponding to a mini-batch, although vertical weight stashing (vertical sync) \cite{narayanan2019pipedream} is employed to maintain version consistency during a forward pass (or backward pass). TiMePReSt introduces a novel approach of intra-batch parallelism to restrict total computation and communication time, and thereby, reducing training time of the DNN under consideration. 
\subsection{Model Architecture}
\begin{figure}[h]
    \centering
    \includegraphics[width=0.5\textwidth, center]{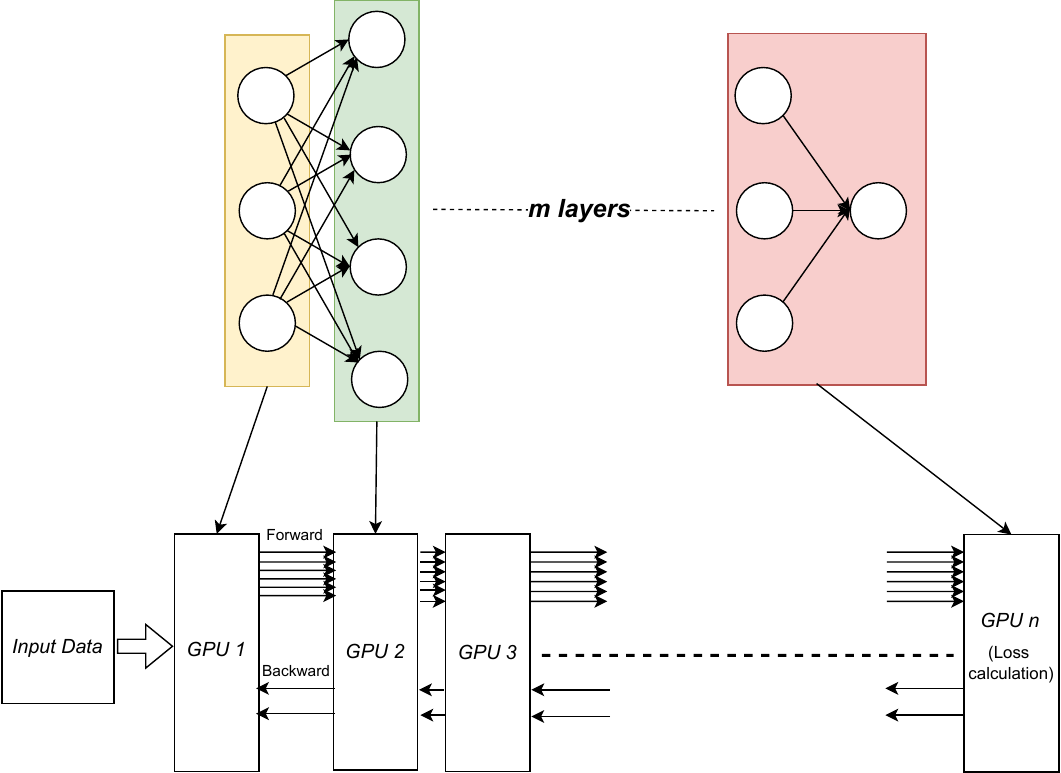}
    \caption{\textbf{A generalized architecture of TiMePReSt having $m-$layered DNN distributed over $n-$sized cluster. Less frequent backward passes than forward passes to incorporate an extra level of intra-batch parallelism. Backward passes start from the GPU where forward passes end. Input data is given to the first GPU only.}}
    \label{fig:Model Architecture}
\end{figure}

\begin{figure}[!ht]
  \begin{subfigure}{1.0\textwidth}
  \centering
    \includegraphics[width=0.99\linewidth]{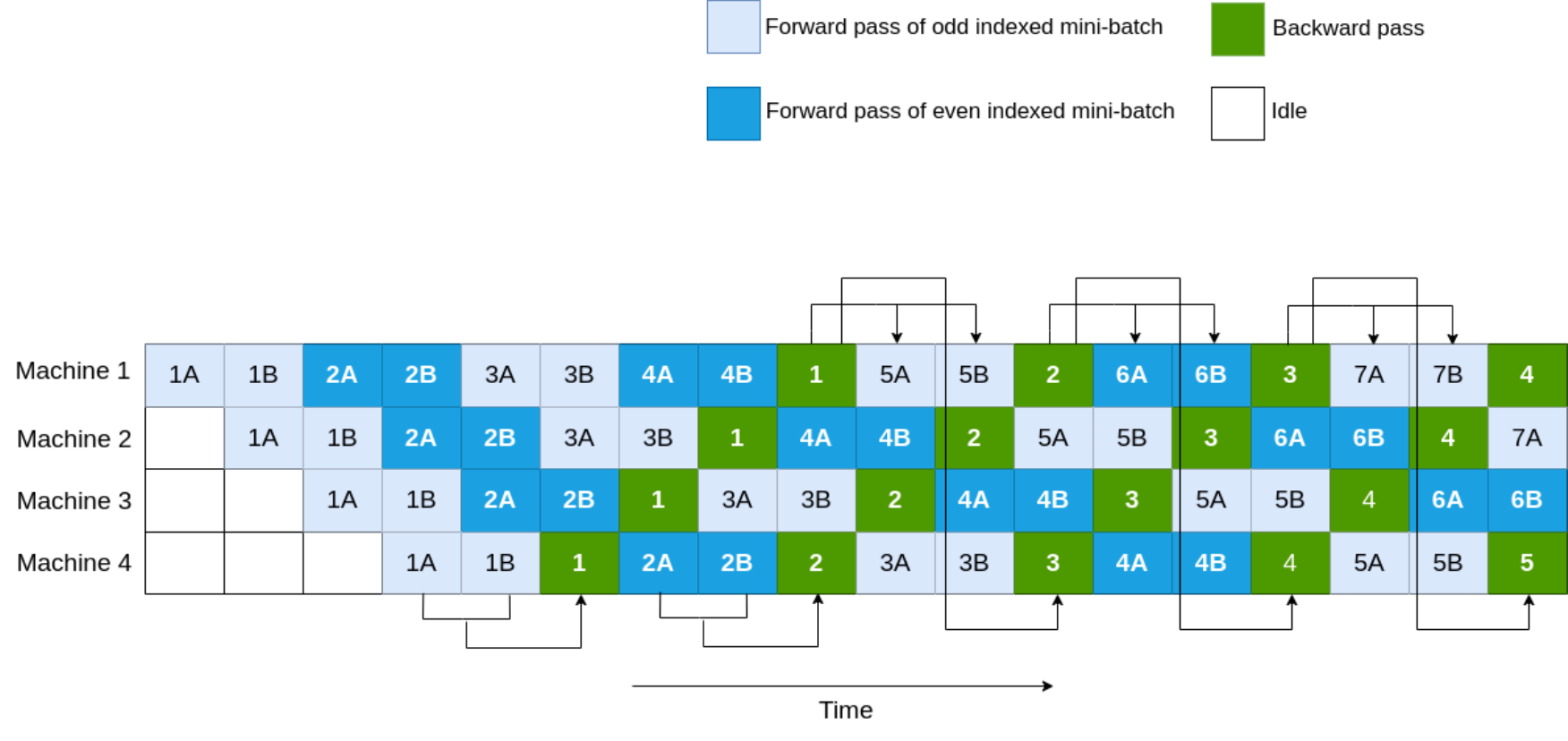}
    \caption{N = 2}
    \label{fig:N = 2}
  \end{subfigure}
  \begin{subfigure}{1.0\textwidth}
  \centering
    \includegraphics[width=1.0\linewidth]{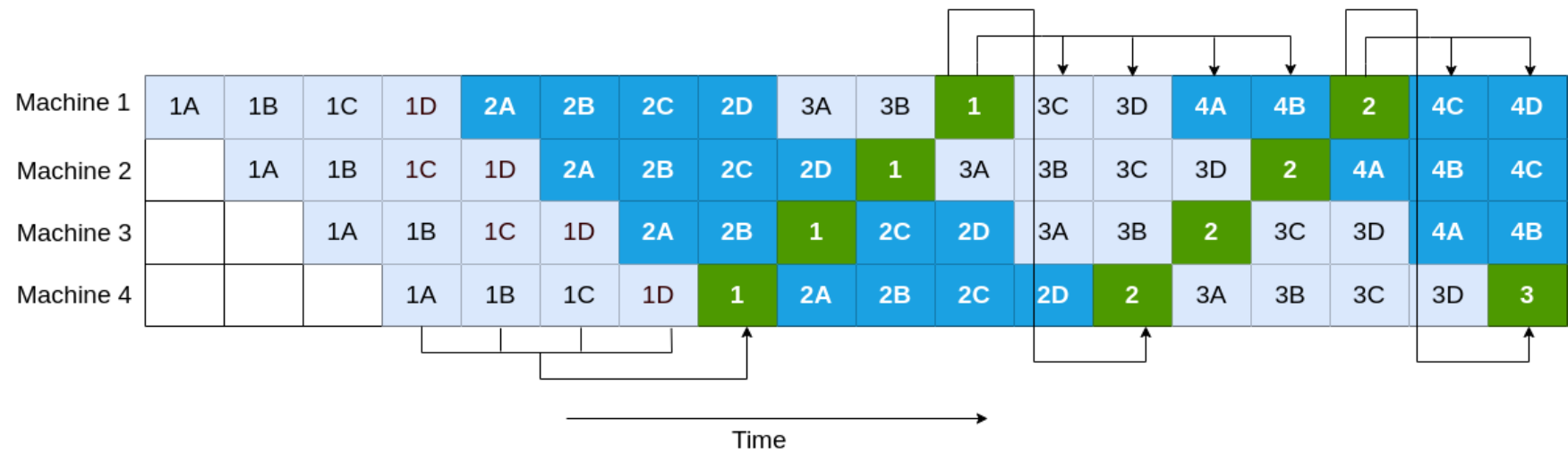}
    \caption{N = 4}
    \label{fig:N = 4}
  \end{subfigure}
  \caption{\textbf{TiMePReSt, the proposed scheme of DNN training using pipeline parallelism for (a) N = 2 and (b) N = 4. An example pipeline parallel training using proposed scheme with four workers is depicted above. Numbers indicate mini-batch ID and alphabets indicate micro-batches. Backward pass occurs less frequently than forward pass.}}
  \label{fig:Proposed scheme of DNN training using pipeline parallelism}
\end{figure}

The backbone of the TiMePReSt's architecture is the pipeline parallel mechanism, where the layers of a DNN are distributed across multiple accelerator (GPU), allocating each set of consecutive layers to an accelerator. In this work, a cluster of two machines, each having a single GPU, has been formed for training purpose. The network is distributed over the cluster so that a balance is maintained among all the member machines in terms of the GPU memory consumption. The machine, which is not responsible for training of the first split of the network, does not need to contain the dataset. Each machine has the information about the sequence of machines through which the training should proceed. Each machine shares the activated output of the last layer of the split with the next machine in the sequence. Once the forward pass is over, prediction error or loss is computed in the last machine of the sequence. The backward pass starts from the same machine, where the forward pass ends. During backward pass, each machine computes gradients of the loss on the weight parameters, and the computed gradients are communicated with the previous machine in the sequence. In order to prevent infinite and circular waiting between forward and backward passes, 
$n$ Forward 1 Backward ($n$F1B) scheduling strategy is introduced in this work, which is a variant of the 1F1B scheduling mechanism used in PipeDream. 
Figure \ref{fig:Model Architecture} shows a generalized architecture of TiMePReSt having m-layered DNN distributed over n-sized cluster.
\\\\In this paper, each mini-batch of size $M$ is split into $N$ smaller micro-batches. Thus, a micro-batch of size $M/N$ becomes the basic data processing unit throughout the pipeline training. Figures \ref{fig:N = 2} and \ref{fig:N = 4} illustrate the workflow of a 4-GPU system with $N = 2$ and $N = 4$, respectively. The number inside each box refers to the index of mini-batches. In case of forward passes, the mini-batch indexes are followed by alphabets, that refer to the micro-batches corresponding to a mini-batch. For better interpretability, three different colors - light blue, dark blue and green are used for forward passes of odd and even indexed mini-batches, and their backward passes respectively; white boxes indicate idle state of GPUs. The forward passes of even and odd indexed mini-batches are denoted by different colors only for better distinction (visually) of consecutive mini-batches. Each mini-batch is trained through the training of $N$ micro-batches. For example, in Figure \ref{fig:N = 2}, micro-batches $1A$ and $1B$ correspond to mini-batch $1$. Similarly, with $N = 4$ (as shown in Figure \ref{fig:N = 4}), micro-batches $1A$ through $1D$ correspond to mini-batch $1$.
\\\\In the next two subsections, we discuss how TiMePReSt tackles the above mentioned challenges in more detail. 
\\\subsection*{\footnotesize Challenge 1: Staleness of Weights}
\label{Challenge 1: Staleness of Weights}
A neural network mimics the way the human brain makes decisions, 
and human always considers the latest updated version of a knowledge base while performing any task. While performing a long-running task, the knowledge base may have been updated since various people can be working on the same version of the knowledge base simultaneously and may discover new information. The earlier version of the knowledge base often looses its significance once updated. The conventional way of training a neural network allows only sequential execution of the mini-batches of a dataset, which is not able to capture this scenario since the weights can be updated only when the currently running mini-batch completes its execution. In other words, a mini-batch can start computation only after the earlier mini-batch finishes. No simultaneous execution of mini-batches or intermediate weight updation is allowed in traditional way of training. However, the pipeline parallelism can relate to the problem since multiple mini-batches start working simultaneously before finishing earlier mini-batches. Thus, weight stashing is a common technique to maintain consistency in weight versions in forward and backward passes, as discussed earlier. As a result, later mini-batches fail to utilize the updated weights computed by the earlier ones if the updates take place during their runtime. This is the problem referred to as staleness of weights.   
\paragraph{Solution: \normalsize\normalfont{ In order to overcome the above mentioned bottleneck of the state-of-the-art pipeline parallelism based methods for training DNNs, we introduce a pipeline-based methodology where each mini-batch backpropagates gradients of the prediction error with respect to the latest updated version of weights rather than the version that was considered during forward propagation (prediction). 
We have already discussed earlier regarding an obvious scenario in pipeline parallelism. During forward propagation of a mini-batch, other mini-batch(s) may complete backpropagation and update the weights. This scenario can be beneficial if and only if the updated weights can be utilized in the upcoming forward and backward passes. The existing pipeline parallelism based DNN training strategies cannot fully utilize the benefit, but TiMePReSt can do it using the proposed strategy. More precisely, TiMePReSt does not allow computing gradients on stale weights} \cite{narayanan2019pipedream}.}
\subsubsection*{\textit{Degree of Staleness}}
In literature, no pipeline parallelism method is found, which is able to overcome staleness problem completely. Unlike existing methods, TiMePReSt is able to achieve zero level of staleness in each pass (forward and backward) while computing gradients and injecting new micro-batches in the pipeline, although a few previous approaches have been able to reduce the extent of staleness to some extent.
One important feature of TiMePReSt is that the $N$ micro-batches corresponding to a mini-batch are allowed to start with different weights while being injected into the pipeline. In Figure \ref{fig:N = 4}, it can be observed that micro-batches C and D corresponding to mini-batch 3 start working with the weights updated by mini-batch 1, which are different from micro-batches A and B. TiMePReSt follows the weight update rule as given below,
\begin{equation}\label{eqn:proposed method}
    \textbf{W}(t+1)=\textbf{W}(t)- \eta \cdot \nabla f\Bigl( \mathbf{W}_1(t - v + 1), \mathbf{W}_2(t - v + 1), \ldots, \mathbf{W}_n(t - v + 1)\Bigr)
\end{equation}
\\We denote $\mathbf{W}_l(t - v + 1)$ as weights of the layers in $l^{th} (l = 1,2,...,n)$ stage after $(t - v + 1)$ mini-batches, $\eta$ as the learning rate, $f$ as the loss function, $\nabla f\Bigl(\mathbf{W}_1(t - v + 1), \mathbf{W}_2(t - v + 1), \ldots, \mathbf{W}_n(t - v + 1)\Bigr)$ as the gradient computed over all samples in the mini-batch, and $\mathbf{W}(t - v + 1)$ is the weights of the entire network (rather than a stage) after $(t - v + 1)$ mini-batches, $v$ as the version difference which refers to the difference between the indices of two mini-batches, one of them access the weight updates performed by the other. The version difference has been discussed in detail in section \ref{Multiple Sequence Problem}.
The update rule shows that TiMePReSt eliminates horizontal weight stashing and computes gradients on the current updated weights, unlike state-of-the-art methods. Thus, it is able to avoid stale weights problem although it allows multiple mini-batches to be active at a time.
\subsubsection*{\textit{Simplicity of Implementation}}
With horizontal weight stashing, gradients in a stage are computed with weights, that are a number of steps delayed. TiMePReSt removes the delay or staleness and observes no significant deterioration in DNN performance. Moreover, the removal of horizontal weight stashing makes the parallelism less complex. For example, in Figure \ref{fig:N = 2}, mini-batches 3 and 4 use weight updates from mini-batches 1 and 2 respectively during backpropagation, although their forward passes start before any update occurs. Similarly, during backpropagation, mini-batch 5 uses updates from mini-batch 3, however, in forward pass the updates from mini-batch 1 are used. In Figure \ref{fig:N = 4}, mini-batches 2 and 3 use the updates from mini-batches 1 and 2 respectively for backward pass. 
\subsubsection*{\textit{GPU Memory Overhead}}
\label{Parameter State}
TiMePReSt stores all trainable parameters associated with the layers assigned to the stage directly in the corresponding GPU memory. It computes gradients on the most recent version of weights once the forward pass is over. When a newer version of the weights becomes available, the previous version is stored until a forward pass, that uses it, is completed. Unlike TiMePReSt, in existing methods, the waiting time before deletion would add the time required for a backward pass also. Thus, in TiMePReSt, GPU memory is released more frequently, which reduces the GPU memory overhead to some extent. 
\subsection*{\footnotesize Challenge 2: High Training Time}
\label{Challenge 2: High Training Time}
As discussed earlier, large sized networks, which cannot be accomodated in a single accelerator, can be distributed across multiple accelerators. Despite the advantage, lack of time-efficiency is another bottleneck of pipeline-parallel distributed DNN training system, which we have addressed in this paper. Communication overhead and frequent parameter updates are two major challenges in the way of achieving desired time-efficiency and they contribute significantly to consume high training time. Cost per communication is directly proportional to the amount of information to be transferred in each communication, 
and it quantity can be adjusted with mini-batch size. We observe that larger mini-batch size increases time required per communication, although less number of communication is required. On the other hand, smaller mini-batch size reduces time required per communication. However, it results in more forward and backward passes, thereby, more number of communications. Moreover, smaller mini-batches cause more frequent parameter updates. As a result, overall training time increases since backward pass is commonly be more computationally expensive and time-consuming than forward pass, and parameter updates are also expensive in terms of time. Thus, maintaining a balance between mini-batch size and training time is important to achieve a time-efficient system. 
\paragraph{Solution: \normalsize\normalfont{In order to maintain a harmony between mini-batch size and training time, TiMePReSt divides a larger mini-batch into smaller micro-batches and performs their forward passes in the pipelining manner. However, their backward passes does not start immediately, unlike conventional way of training and asynchronous pipeline parallelism techniques. Once all the micro-batches corresponding to a mini-batch complete their forward passes, the backward pass starts for the mini-batch considering the average prediction error or loss of all the micro-batches. For example, in Figure \ref{fig:N = 2}, mini-batch 1 is divided into two micro-batches namely 1A and 1B. The backward pass of mini-batch 1 starts once the forward passes of both 1A and 1B complete. It happens for the other mini-batches too. The proposed strategy ensures getting effect of a mini-batch without processing entire mini-batch at a time.}}
\subsubsection*{\textit{Overlapping of Computation and Communication}}
\begin{figure}[ht]
    \centering
    \includegraphics[width=0.5\textwidth, center]{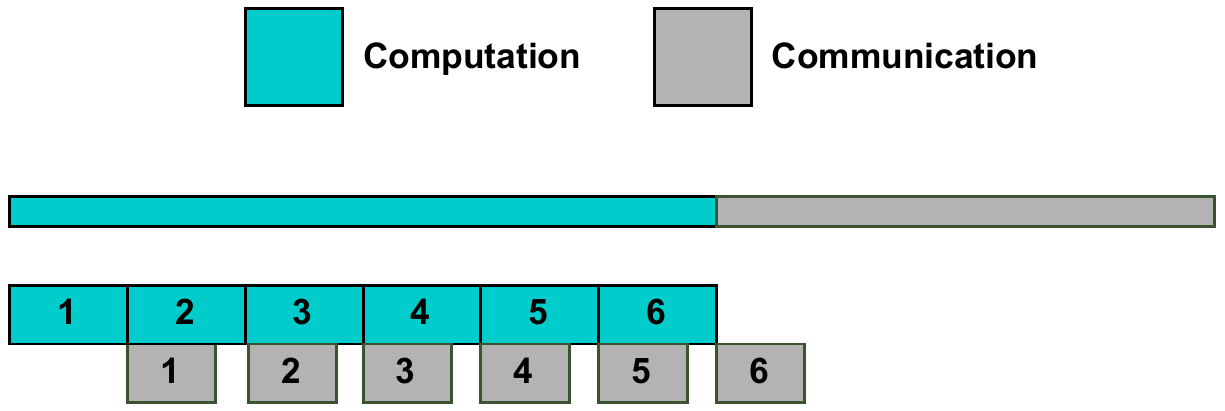}
    \caption{\textbf{Proposed intra-batch parallelism strategy. 
    Overlapping of computation and communication using proposed scheme reduces training time significantly.}}
    \label{fig:comp_comm_overlap}
\end{figure}

GPUs compute and communicate synchronously. That is, the outputs of the computation should be sent after the entire computation has been completed \cite{zhao2022bapipe}. Thus, smaller micro-batches enable GPUs to perform computation and communication of the micro-batches in parallel, since GPUs consider them as individual mini-batches during forward pass. In this way, \textit{intra-batch parallelism} is employed in our method to incorporate an extra level of parallelism. Thus, overall training time is reduced. Figure \ref{fig:comp_comm_overlap} provides a clear visualisation of the scenario. For example, in Figure \ref{fig:N = 2}, micro-batch 1A is executing forward pass on machine 2 while 1B is executing that on machine 1. If mini-batch 1 would not be decomposed, the system would have to wait for completion of stage 1 for all the data samples corresponding to mini-batch 1 before initiating stage 2.
\subsubsection*{\textit{Computation and Communication Time Vs Mini-batch Size}}Executing forward pass on smaller micro-batches reduces cost per communication, which is the benefit of having smaller mini-batches. However, number of communication increases with downsizing mini-batches. The proposed approach helps to restrict number of communications, by limiting the number of backward passes, since it is both time and computation expensive. In other words, number of communication increases during forward passes only, but not in backward pass. Less number of communications could be achieved using larger mini-batch size, although cost per communication would increase drastically. TiMePReSt combines the advantages of both smaller and larger mini-batches. More precisely, both computation and communication costs decrease without compromising with mini-batch size. 
\subsubsection*{\textit{Efficient Memory Usage}}In traditional pipelining technique, the maximum size of each mini-batch depends on the individual GPU memory capacities of the worker machines. In the proposed method, although the mini-batch size is larger, however, due to smaller micro-batches, entire mini-batch need not reside in GPU memory at a time. It requires only to accomodate a single micro-batch at a time during forward pass, although the backward pass gets the effect of entire mini-batch through the average loss of all the micro-batches, rather than a single micro-batch. Thus, we can consider larger sized mini-batch compared to traditional pipelining technique due to efficient memory usage. For example, in Figure \ref{fig:N = 2} and \ref{fig:N = 4}, if the micro-batch size is $n$, a DNN is trained on $2n$ and $4n$ samples respectively in a single mini-batch. However, only $n$ samples participate at a time during forward pass although the gradient is computed on the collective prediction error of all $2n$ and $4n$ samples.
\subsection{Work Scheduling}
\label{Work Scheduling}
TiMePReSt is a bi-directional pipeline, where an input mini-batch executes forward pass through the pipeline stages and then backward pass through the same pipeline stages in reverse order, starting from the machine where forward pass ends. Each active mini-batch is divided into some equal-sized micro-batches. Each of them may be in a different stage of forward pass at a time. Once all of them complete forward passes, a backward pass starts gradient computation on their average loss. The experiments show that TiMePReSt helps to complete similar sized mini-batch in less time than PipeDream \cite{narayanan2019pipedream}. The workers in the cluster often need to 
make a choice between performing forward pass of a micro-batch or backward pass of a mini-batch. We introduce $n$F1B scheduling to maintain a balanced execution flow of forward and backward passes.
\\\\Once the forward passes of $n$ micro-batches corresponding to the first mini-batch complete, each worker starts performing single backward pass. As a result, a chain is formed for each worker consisting of repeated `$n$ forward passes followed by 1 backward pass'. Since we have more number of forward passes than backward passes, backward pass gets priority on arrival to prevent waiting for many forward passes. For example, in Figures \ref{fig:N = 2} and \ref{fig:N = 4}, we observe the backward passes are performed at each third and fifth time points respectively after first backward pass finishes, since each mini-batch is divided into two and four micro-batches respectively. 
\subsection{Checkpointing}
\label{Checkpointing}
Fault tolerance is a major concern since large DNNs require a number of hours to be trained. Similar to PipeDream\cite{narayanan2019pipedream}, TiMePReSt allows periodic stage-wise checkpointing of model parameters for fault tolerance at the end of every epoch. Each stage saves its model parameters locally after the backward pass for the last mini-batch in an epoch. Thus, checkpointing does not require communication to other worker machines. Periodic checkpointing enables the system to start from the most recent checkpoints for all stages, if the training resumes after a sudden termination of a run. In case of a stage failure, checkpointing enables the system to resume the training process from the last successfully completed epoch. Thus, periodic checkpointing in all stages makes the system fault tolerant in case of abnormal termination of the training. Since the checkpointing is performed independently in all the stages, the system also becomes tolerant of stage failure.
\subsection{Multiple Sequence Problem}\label{Multiple Sequence Problem}
In Figure \ref{fig:N = 2}, we can see that, the effect of the weight updation using mini-batch $1$ is propagating through mini-batches $3,5$ and $7$. Simultaneously, weight updation using mini-batch $2$ is propagating through mini-batches $4$ and $6$. We consider the sequence of backward passes only, for the ease of constructing sequences of mini-batches. The grouping of the mini-batches remains the same even if both forward and backward passes are considered. However, it becomes difficult to construct the sequences of mini-batches through which the effect of weight updation is propagated. In Figure \ref{fig:N = 2}, we can see that two distinct sequences of mini-batches are running in parallel, which are getting no interaction with each other till the end of an epoch. Thus, the ultimately learnt weights are the results of the cumulative efforts of a few mini-batches only, rather than all. This scenario may weaken the training process of a DNN.
\\\\In Figure \ref{fig:N = 4}, no such discrepancy occurs. The effect of the weight updation using mini-batch $1$ is propagating through mini-batches $2,3,4$ and so on. Thus, the ultimately learnt weights are the results of the cumulative efforts of all the mini-batches. 
\subsection*{\textit{Solution}}
The reason behind not occurring the above mentioned multiple sequence problem in Figure \ref{fig:N = 4} is the version difference $(v)$. The version difference $(v)$ refers to the difference between the indices of two consecutive mini-batches which participate in the same sequence. For example, in Figure \ref{fig:N = 2}, $v = 2$ for both the sequences $\{1,3,5,7,....\}$ and $\{2,4,6,.....\}$. In contrast, only a single sequence is there in Figure \ref{fig:N = 4}, which is having $v = 1$. It is clearly observed that the $v$ should have value of unity preferably, in order to address the above mentioned multiple sequence problem. In this paper, $v$ depends on the number of workers $(W)$ and number of micro-batches $(N)$, which motivates us to find out the mathematical relationship between $W$ and $N$. The domains of $W$, $N$ and $v$ are given below.
\begin{equation}\label{eqn:domain of W}
    W \in \{k | k \in \mathbb{Z}^{+} \wedge k \geq 2\}
\end{equation}
\begin{equation}\label{eqn:domain of N}
    N \in \{k | k \in \mathbb{Z}^{+} \wedge k \geq 2\}
\end{equation}
\begin{equation}\label{eqn:domain of v}
    v \in \{k | k \in \mathbb{Z}^{+} \wedge k \geq 1\}
\end{equation}
In this paper, we prefer to set $W$ and $N$ values in a certain way so that $v = 1$.

\subsection*{Mathematical Relationship between $W$ and $N$}
Here, the number of time-points required for the forward passes of mini-batches $1$ and $2$ are denoted by $f_1$ and $f_2$ respectively. We observe in Figures \ref{fig:N = 2} and \ref{fig:N = 4} that $f_1$ and $f_2$ can be represented mathematically as
\begin{equation}\label{eqn:FP_1}
    f_1 = W + N - 1
\end{equation}
\begin{equation}\label{eqn:FP_2}
    \begin{split}
     f_2 &= f_1 + 1 \\
             &= W + N
    \end{split}
\end{equation}
We observe that each index-wise successive mini-batch takes $1$ more time-point to complete its forward pass than the preceding one. Thus, any two index-wise consecutive mini-batches can be considered rather than mini-batches $1$ and $2$. The number of time-points required for the backward pass of a mini-batch can be represented as
\begin{equation}\label{eqn:BP}
    b = W
\end{equation}
We observe that $v$ can have value of unity if and only if the backward pass using mini-batch $2$ does not start before that using mini-batch $1$ ends. In other words, mini-batch $2$ must have to continue its forward pass until mini-batch $1$ completes its backward pass. There should not be any overlapping between their backward passes. The mathematical representation of this condition can be written as
\begin{equation}\label{eqn:v=1 condition basic}
    f_1 + b - N \leq f_2,\text{ iff } v = 1
\end{equation}
\begin{equation}\label{eqn:v>1 condition basic}
    f_1 + b - N > f_2,\text{ iff } v > 1
\end{equation}
\\In the above inequation \ref{eqn:v=1 condition basic}, $(f_1 + b - N)$ represents the minimum number of time-points the forward pass of mini-batch 2 should continue. By replacing $f_1$, $f_2$ and $b$ in inequation \ref{eqn:v=1 condition basic} with Equations \ref{eqn:FP_1}, \ref{eqn:FP_2}, and \ref{eqn:BP} respectively, we get
\begin{equation}\label{eqn:v=1 condition derived}
    \begin{split}
        (W + N - 1) + W - N \leq W + N \\
        (2\times W) - 1  \leq W + N \\
        W \leq N + 1,\text{ iff } v = 1
    \end{split}
\end{equation}
\\Similarly, from inequation \ref{eqn:v>1 condition basic}, we will get
\begin{equation}\label{eqn:v>1 condition derived}
    W > N + 1,\text{ iff } v > 1 
\end{equation}
\subsection*{Mathematical Expression for Version Difference}
In this paper, we are interested merely in the scenario of $v = 1$. However, the research can be extended to the scenario $v > 1$ also. Experiments can be conducted to execute a comparative study on the model performances for different values of $v$. Thus, we are motivated to introduce a mathematical expression for $v$ which shows its relationship with $W$ and $N$, so that the values of $W$ and $N$ need not be set experimentally. Additionally, $v$ can be computed for a given $W$ and $N$, or vice-versa, without drawing diagrams like Figures \ref{fig:N = 2} and \ref{fig:N = 4}. 
We observe that $v > 1$ if and only if the backward passes of at least two mini-batches overlap during training. In other words, it is mandatory to ensure the execution of the backward passes of different mini-batches one after another, in order to maintain $v = 1$. No overlapping of the backward passes is ensured if and only if the forward pass of a mini-batch is not complete before the backward pass of the preceding mini-batch is over, since a backward pass of a mini-batch starts immediately after its forward pass ends. We can observe in Figures \ref{fig:N = 4}, \ref{fig:N = 2, W = 3}, and Figures \ref{fig:N = 2}, \ref{fig:N = 2, W = 5} that the above mentioned scenario for the overlapping of the backward passes arises if and only if $W > N + 1$ holds and more than $2$ mini-batches enter into the pipeline with all its micro-batches before the forward pass of the first mini-batch ends. This number of mini-batches can be expressed as $\frac{f_1}{N}$, which equals to $\floor*{\frac{W + N - 1}{N}}$ , $N$ being the number of time-points required for a mini-batch to enter into the pipeline with all its micro-batches. The floor($\left\lfloor  \right\rfloor$) operator helps to exclude any partially entered mini-batch from this particular computation. Thus, the above observation can be summarized as - the increment of $\floor*{\frac{W + N - 1}{N}}$ results in the increment of $v$. 
\begin{figure}[!ht]
  \begin{subfigure}{1.0\textwidth}
  \centering
    \includegraphics[width=0.99\linewidth]{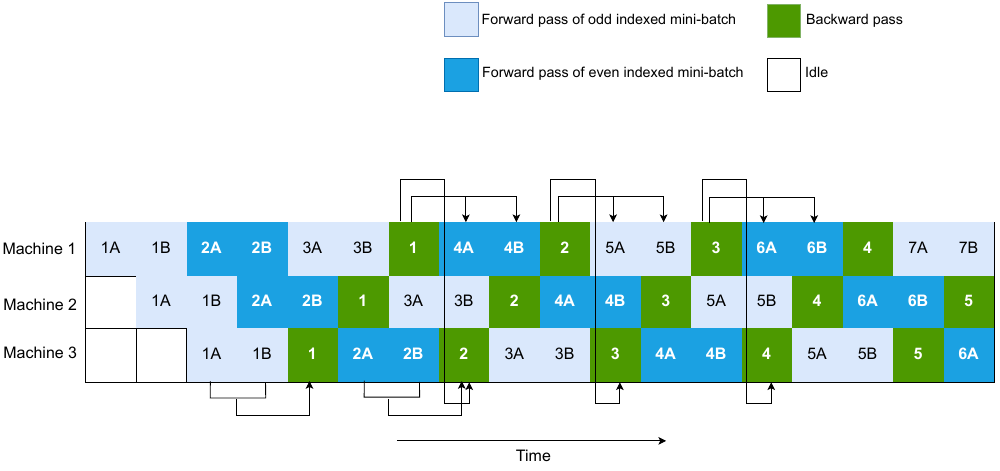}
    \caption{W = 3}
    \label{fig:N = 2, W = 3}
  \end{subfigure}
  \begin{subfigure}{1.0\textwidth}
  \centering
    \includegraphics[width=1.0\linewidth]{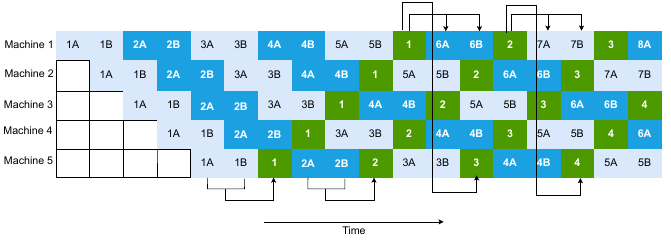}
    \caption{W = 5}
    \label{fig:N = 2, W = 5}
  \end{subfigure}
  \caption{\textbf{TiMePReSt, the proposed scheme of DNN training using pipeline parallelism for (a) W = 3 and (b) W = 5. An example pipeline parallel training using proposed scheme is depicted above which is having each mini-batch divided into three micro-batches. Numbers indicate mini-batch ID and alphabets indicate micro-batches. Backward pass occurs less frequently than forward pass.}}
  \label{fig:Proposed scheme with W=3,5 and N = 2 of DNN training using pipeline parallelism}
\end{figure}
\subsection*{Case $1: v = 1$}
From Equation \ref{eqn:v=1 condition derived}, we get
\begin{equation}
    \begin{split}
        W \leq N + 1 \text{ iff } v = 1 \\
        W + N - 1 \leq 2\times N \\
        \frac{W + N - 1}{N} \leq 2 
    \end{split}
\end{equation}
We know that the next mini-batch starts entering into the pipeline during the forward pass of the earlier mini-batch. Thus, $\frac{W + N - 1}{N}$ cannot be equal to $1$.
Thus, we can say that,
\begin{equation}\label{eqn:number of mini-batch in one forward pass without floor}
    1 < \frac{W + N - 1}{N} \leq 2
\end{equation}
Since, $\floor*{\frac{W + N - 1}{N}}$ cannot be less than $1$, we can write
\begin{equation}\label{eqn:number of mini-batch in one forward pass}
    \floor*{\frac{W + N - 1}{N}} \in \{k | k \in \mathbb{Z}^{+} \wedge k \leq 2\}
\end{equation}
Thus, we can summarize the above discussion like below:
\begin{equation}\label{eqn:v=1 condition derived new}
    \floor*{\frac{W + N - 1}{N}} \in \{k | k \in \mathbb{Z}^{+} \wedge k \leq 2\} \text{ iff } v = 1 
\end{equation}
By comparing the values of $v$ and $\floor*{\frac{W + N - 1}{N}}$, we have 
\begin{equation}\label{eqn:relationship between v and (W+N-1)/N for v = 1}
    \begin{split}
        v \leq \floor*{\frac{W + N - 1}{N}} \\
        v \leq \frac{W + N - 1}{N} \\        
        v = \frac{W + N - 1}{N} - x \quad 0\leq x \leq 1
    \end{split}
\end{equation}
Since $0\leq x\leq 1$, and we observe that for a given $W$, $x$ varies with $N$. If $N$ increases, $\frac{W + N - 1}{N}$ decreases. Resultantly, $x$ decreases as $v = 1$. Similarly, $x$ increases with $N$ decreases. Thus, we assume $x\sim \frac{1}{N}$. By replacing $x$ in Equation \ref{eqn:relationship between v and (W+N-1)/N for v = 1} we get,
\begin{equation}\label{eqn:intermediate expression of v}
    \begin{split}
        v \approx \frac{W + N - 2}{N} \\
    \end{split}
\end{equation}
\subsubsection*{\textit{Case 1a}}
If $\frac{W + N - 1}{N} = 2$, then 
\begin{equation}\label{eqn:decide floor case 1}
    \begin{split}
        1 < \frac{W + N - 2}{N} < 2, \quad \text{ as } 0 < \frac{1}{N} \leq 0.5 \\
        \floor*{\frac{W + N - 2}{N}} = 1 \\
    \end{split}
\end{equation}
Since $v = 1$ as per our assumption, we have 
\begin{equation}\label{eqn:final expression of v case 1}
    \begin{split}
        v = \floor*{\frac{W + N - 2}{N}}
    \end{split}
\end{equation}
\subsubsection*{\textit{Case 1b}}
If $1 < \frac{W + N - 1}{N} < 2$, then $\frac{W + N - 1}{N}$ achieves its minimum value when $W = 2$. If $W = 2$, then
\begin{equation}\label{eqn:final expression of v case 2}
    \begin{split}
        \frac{W + N - 1}{N} = 1 + \frac{1}{N} \\
        \frac{W + N - 1}{N} - \frac{1}{N} = 1 \\
        \floor*{\frac{W + N - 1}{N} - \frac{1}{N}} = 1 \\
        \floor*{\frac{W + N - 2}{N}} = v \quad \text{ for } v = 1
    \end{split}
\end{equation} 
\subsection*{Case $2: v > 1$}
\begin{figure}[ht]
    \centering
    \includegraphics[width=1\textwidth, center]{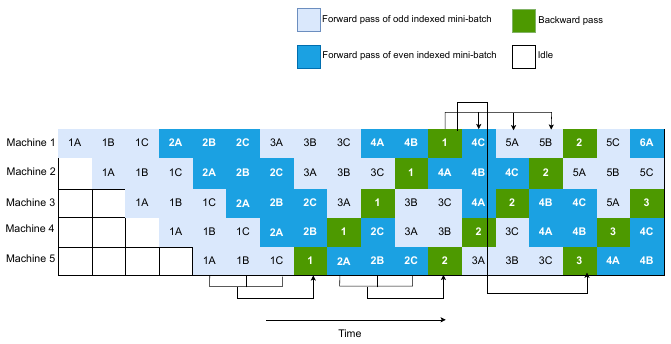}
    \caption{\textbf{TiMePReSt, the proposed scheme of DNN training using pipeline parallelism for W = 5 and N = 3. An example pipeline parallel training using proposed scheme is depicted above which is having each mini-batch divided into three micro-batches. Numbers indicate mini-batch ID and alphabets indicate micro-batches. Backward pass occurs less frequently than forward pass.}}
    \label{fig:N = 3, W = 5}
\end{figure}
From Equation \ref{eqn:v>1 condition derived} we get
\begin{equation}
    \begin{split}
        W > N + 1 \iff v > 1 \\
        W + N - 1 > 2\times N \\
        \frac{W + N - 1}{N} > 2 
    \end{split}
\end{equation}
Thus, we can say 
\begin{equation}\label{eqn:v>1 condition derived new}
    \floor*{\frac{W + N - 1}{N}} \in \{k | k \in \mathbb{Z}^{+} \wedge k > 2\} \iff v > 1 
\end{equation}
As depicted by Figures \ref{fig:N = 2, W = 5} and \ref{fig:N = 3, W = 5}, for any combination of N and W, which satisfies the relation $W > N + 1$, we can say
\begin{equation}\label{eqn:relationship between v and (W+N-1)/N for v > 1}
    \begin{split}
        v \leq \floor*{\frac{W + N - 1}{N}} \\
        v \leq \frac{W + N - 1}{N} \\        
    \end{split}
\end{equation}
Since, equations \ref{eqn:relationship between v and (W+N-1)/N for v = 1} and \ref{eqn:relationship between v and (W+N-1)/N for v > 1} are the same, we can say that equation \ref{eqn:intermediate expression of v} is true for $v > 1$ also. We can write $W > N + 1$ and $W \geq N + 2$ interchangeably. For $W \geq N + 2$,

\begin{equation}\label{eqn:final expression of v when v > 1}
    \begin{split}
        \frac{W + N - 2}{N} \geq 2 \\
        \floor*{\frac{W + N - 2}{N}} > 1 \\
        \floor*{\frac{W + N - 2}{N}} = v, \quad \text{ for } v > 1
    \end{split}
\end{equation}

\section{Evaluation}
\label{Evaluation}
\begin{figure}[h]
  \begin{subfigure}{.31\textwidth}
  \centering
    \includegraphics[width=1.0\linewidth]{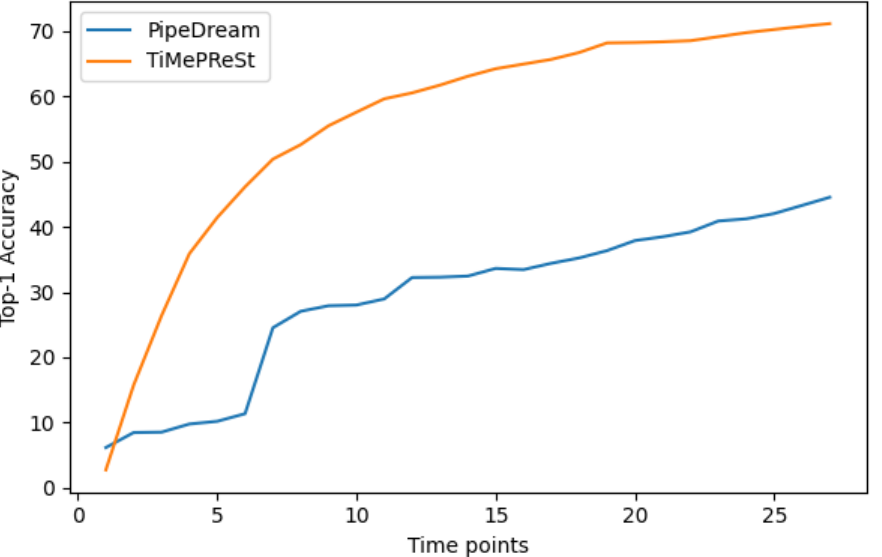}
    \caption{Top-1 accuracy to time}
    \label{fig:Top-1 accuracy to time_cifar100}
  \end{subfigure}
  \begin{subfigure}{.31\textwidth}
  \centering
    \includegraphics[width=1.0\linewidth]{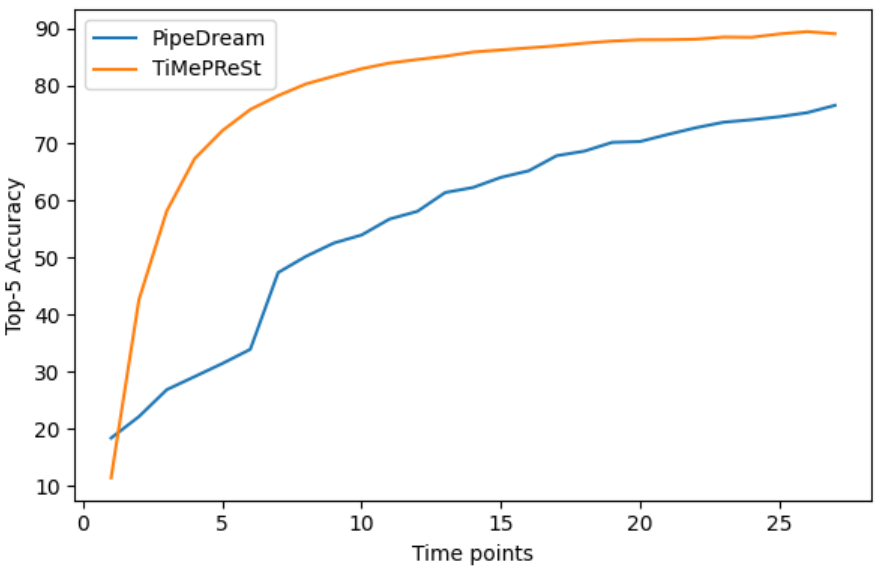}
    \caption{Top-5 accuracy to time}
    \label{fig:Top-5 accuracy to time_cifar100}
  \end{subfigure}
  \begin{subfigure}{.33\textwidth}
  \centering
    \includegraphics[width=0.99\linewidth]{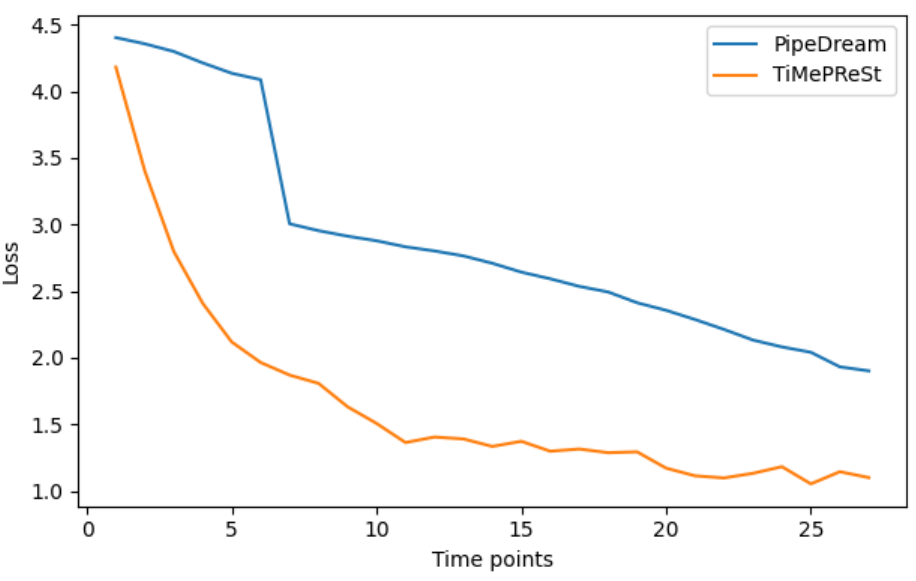}
    \caption{Loss comparison}
    \label{fig:Loss comparison_time_cifar100}
  \end{subfigure}
  \caption{\textbf{Performance comparison of TiMePReSt and PipeDream (VGG-16 on CIFAR-100)}}
  \label{fig:Performance Comparison of Proposed Method and PipeDream (VGG16 on CIFAR-100)}
\end{figure}

\begin{figure}[h]
  \begin{subfigure}{.31\textwidth}
  \centering
    \includegraphics[width=0.99\linewidth]{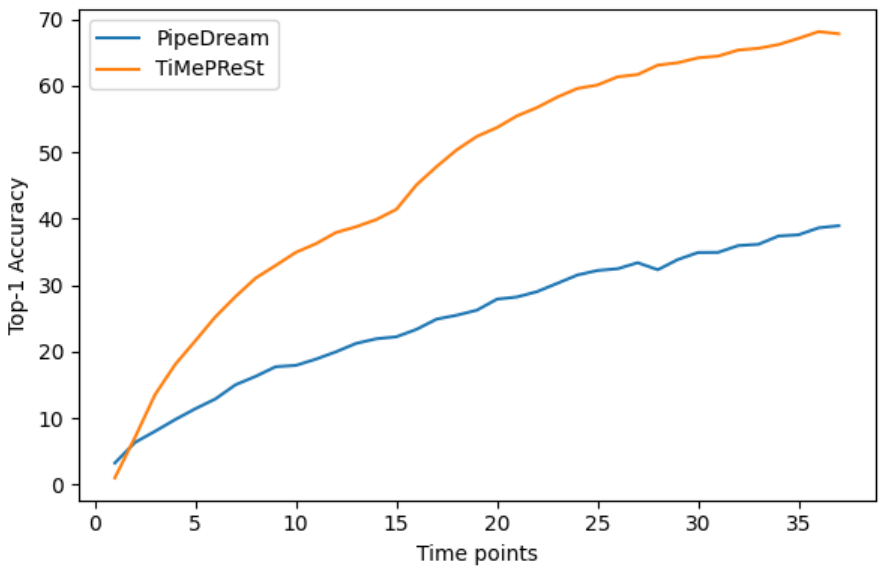}
    \caption{Top-1 accuracy to time}
    \label{fig:Top-1 accuracy to time_tiny_imagenet}
  \end{subfigure}
  \begin{subfigure}{.31\textwidth}
  \centering
    \includegraphics[width=1.0\linewidth]{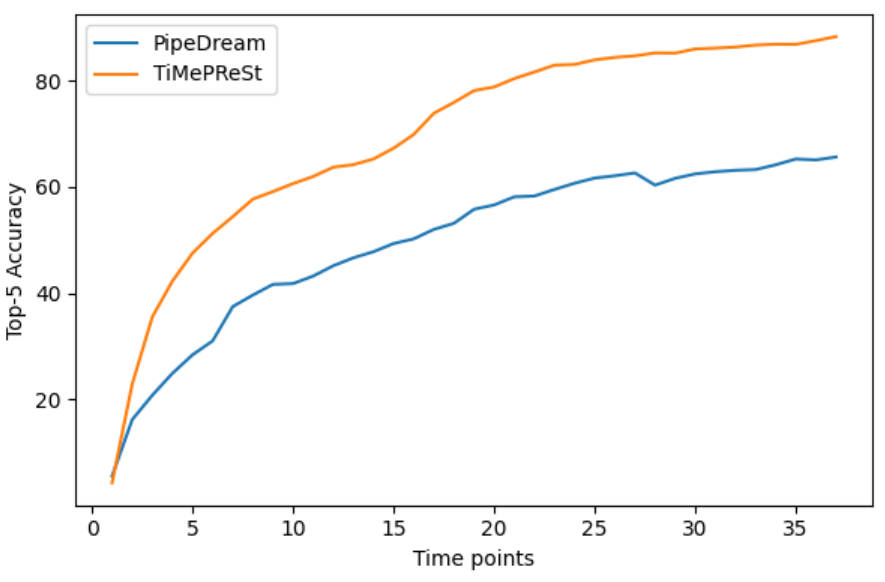}
    \caption{Top-5 accuracy to time}
    \label{fig:Top-5 accuracy to time_tiny_imagenet}
  \end{subfigure}
  \begin{subfigure}{.33\textwidth}
  \centering
    \includegraphics[width=0.99\linewidth]{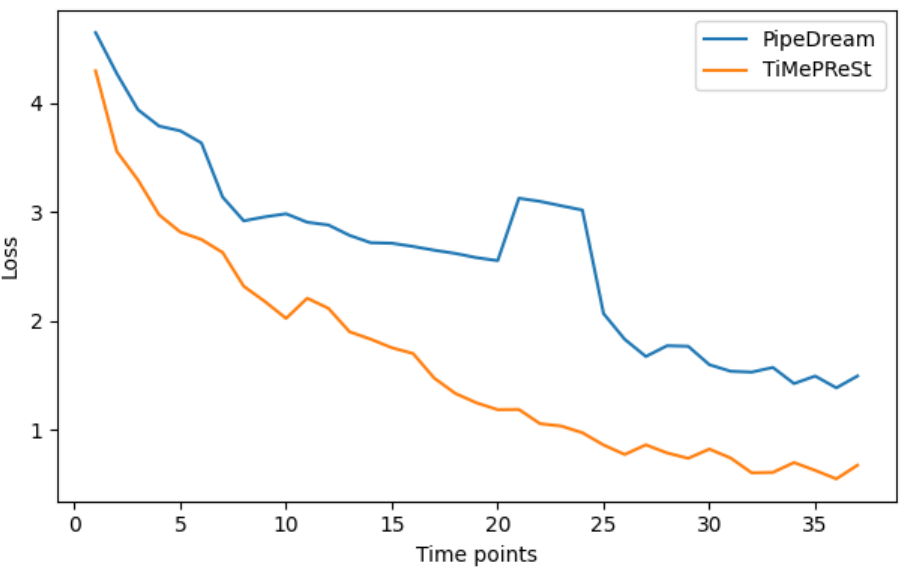}
    \caption{Loss comparison}
    \label{fig:Loss comparison_time_tiny_imagenet}
  \end{subfigure}
  \caption{\textbf{Performance comparison of TiMePReSt and PipeDream (VGG16 on Tiny-ImageNet-200)}}
  \label{fig:Performance Comparison of Proposed Method and PipeDream (Tiny_imagenet)}
\end{figure}

\begin{figure}
  \centering
  \begin{subfigure}{.48\textwidth} 
    \centering
    \includegraphics[width=\linewidth]{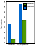}
    \caption{Hardware Efficiency (Time per epoch)}
    \label{fig:Time per epoch}
  \end{subfigure}%
  \hfill
  \begin{subfigure}{.48\textwidth}
    \centering
    \includegraphics[width=\linewidth]{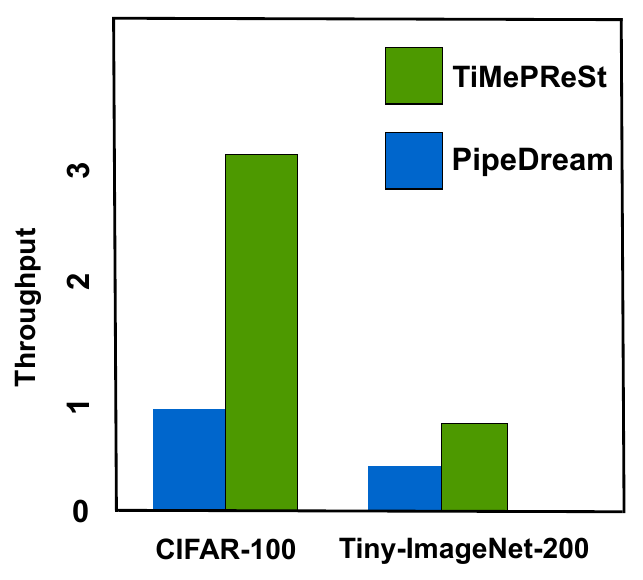}
    \caption{Throughput (epochs/hr)}
    \label{fig:Throughput}
  \end{subfigure}
  \caption{\textbf{Comparison between TiMePReSt and PipeDream, based on hardware efficiency and throughput of training VGG-16 on CIFAR-100 and Tiny-ImageNet-200 image classification datasets.}}
  \label{fig:Comparison between TiMePReSt and PipeDream, based on hardware efficiency and throughput}
\end{figure}

\paragraph{\textit{Time needed to achieve target accuracy:} \normalsize\normalfont{We compare TiMePReSt and PipeDream time-to-accuracy for VGG-16 on CIFAR-100 and Tiny-ImageNet-200 image classification datasets using a cluster consisting of two machines having single GPU each. One is NVIDIA Quadro RTX 6000 with 24 GB of GPU memory, another is NVIDIA GeForce RTX 2080 with 12 GB of GPU memory. Figures \ref{fig:Top-1 accuracy to time_cifar100} and \ref{fig:Top-5 accuracy to time_cifar100} show that TiMePReSt reaches target top-1 and top-5 accuracy %
much faster than PipeDream respectively, in case of CIFAR-100, whereas Figures \ref{fig:Top-1 accuracy to time_tiny_imagenet} and \ref{fig:Top-5 accuracy to time_tiny_imagenet} show similar results for Tiny-ImageNet-200. With no compromise with mini-batch size, introducing extra level of intra-batch parallelism in forward pass and limiting the frequency of backward passes are the main factors behind achieving the time-efficiency over PipeDream.}}
\paragraph{\textit{Accuracy reached after equal execution time:} \normalsize\normalfont{TiMePReSt achieves better top-1 and top-5 accuracies 
higher than PipeDream, after equal training time of VGG-16 on both CIFAR-100 and Tiny-ImageNet-200. Figures \ref{fig:Top-1 accuracy to time_cifar100}, \ref{fig:Top-5 accuracy to time_cifar100}; and Figures \ref{fig:Top-1 accuracy to time_tiny_imagenet}, \ref{fig:Top-5 accuracy to time_tiny_imagenet} show this comparison as the VGG-16 network is trained over time.}}
\paragraph{\textit{Reduction of loss after equal execution time:} \normalsize\normalfont{Figures \ref{fig:Loss comparison_time_cifar100} and \ref{fig:Loss comparison_time_tiny_imagenet} show that TiMePReSt is able to reduce the loss 
to a better extent than where it starts from, whereas PipeDream achieves less reduction than TiMePReSt, after similar execution time of VGG-16 on CIFAR-100 and Tiny-ImageNet-200 respectively. In other words, TiMePReSt is able to reduce the loss much faster than PipeDream.}}
\paragraph{\textit{Number of epochs per hour:} \normalsize\normalfont{The mini-batch size is taken differently for different-sized datasets, although for each dataset, both TiMePReSt and PipeDream consider the same mini-batch size for fair comparison. TiMePReSt performs more number of epochs than PipeDream in unit time, for both CIFAR-100 and Tiny-ImageNet-200 datasets. In other words, TiMePReSt is more efficient than PipeDream in terms of number of epochs per hour, which is also called \textbf{throughput}\cite{narayanan2019pipedream}. Thus, for better comparability, all plots in Figures \ref{fig:Performance Comparison of Proposed Method and PipeDream (VGG16 on CIFAR-100)} and \ref{fig:Performance Comparison of Proposed Method and PipeDream (Tiny_imagenet)} are having time points in x-axis rather than epochs or actual clock-time. Time interval between two consecutive time points is the single epoch time for any one of the models - PipeDream or TiMePReSt, which one is greater. Thus, the length of a single time-interval varies with datasets, since epoch time varies with datasets for both the models. Figure \ref{fig:Throughput} clearly shows the efficiency of TiMePReSt over PipeDream in terms of throughput, for both the datasets.}}
\begin{figure}[h]
  \begin{subfigure}{.31\textwidth}
  \centering
    \includegraphics[width=0.99\linewidth]{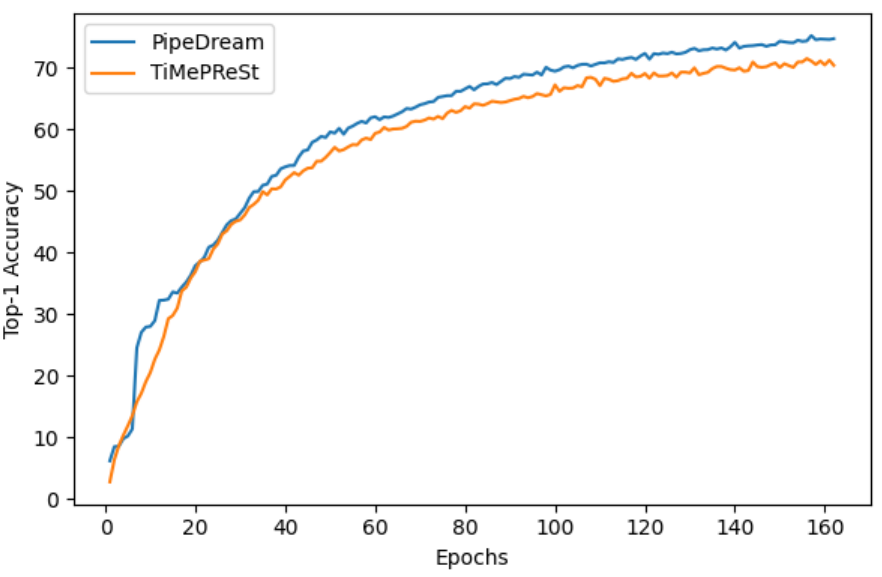}
    \caption{Top-1 accuracy to epoch}
    \label{fig:Top-1 accuracy to epoch_cifar100}
  \end{subfigure}
  \begin{subfigure}{.31\textwidth}
  \centering
    \includegraphics[width=1.0\linewidth]{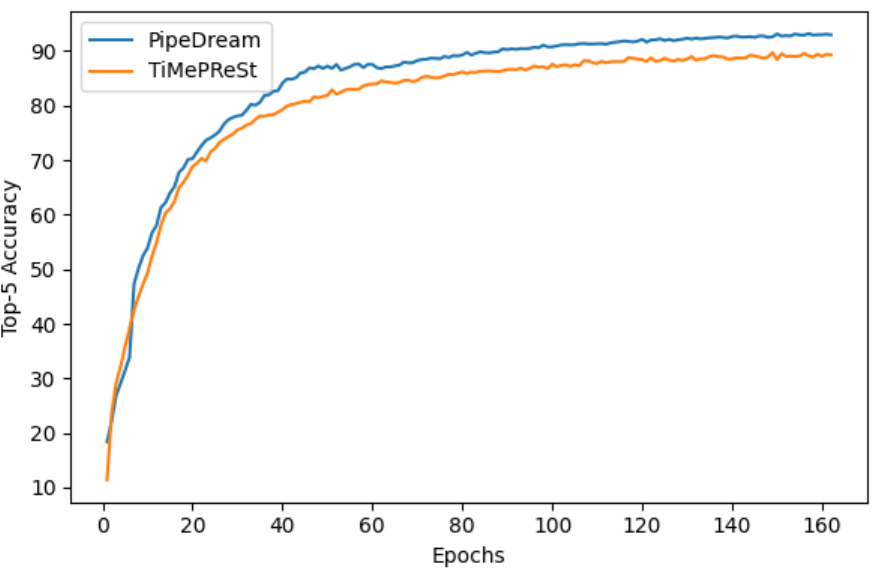}
    \caption{Top-5 accuracy to epoch}
    \label{fig:Top-5 accuracy to epoch_cifar100}
  \end{subfigure}
  \begin{subfigure}{.33\textwidth}
  \centering
    \includegraphics[width=0.99\linewidth]{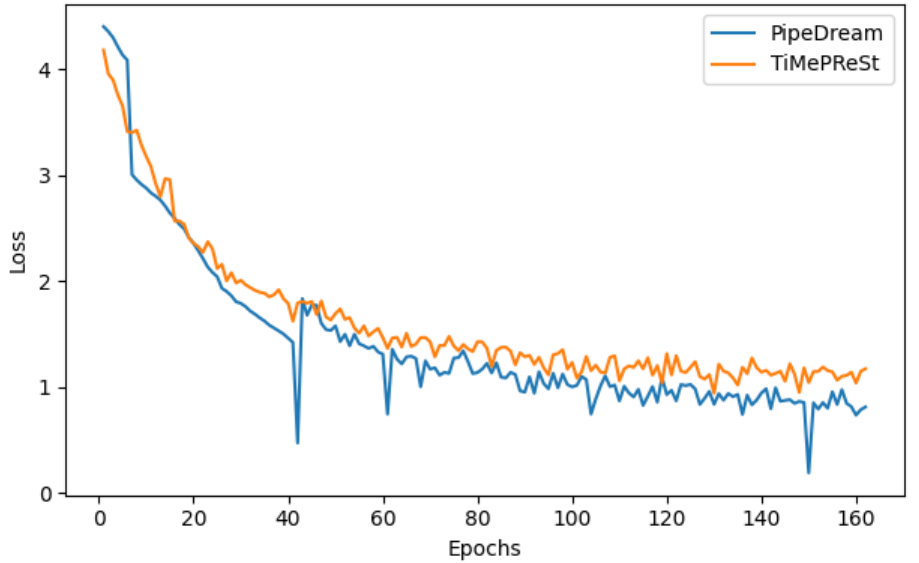}
    \caption{Loss comparison}
    \label{fig:Loss comparison_epoch_cifar100}
  \end{subfigure}
  \caption{\textbf{Statistical efficiency comparison of TiMePReSt and PipeDream (VGG-16 on CIFAR-100)}}
  \label{fig:Statistical Efficiency Comparison of Proposed Method and PipeDream (VGG16 on CIFAR-100)}
\end{figure}

\begin{figure}[h]
  \begin{subfigure}{.31\textwidth}
  \centering
    \includegraphics[width=0.99\linewidth]{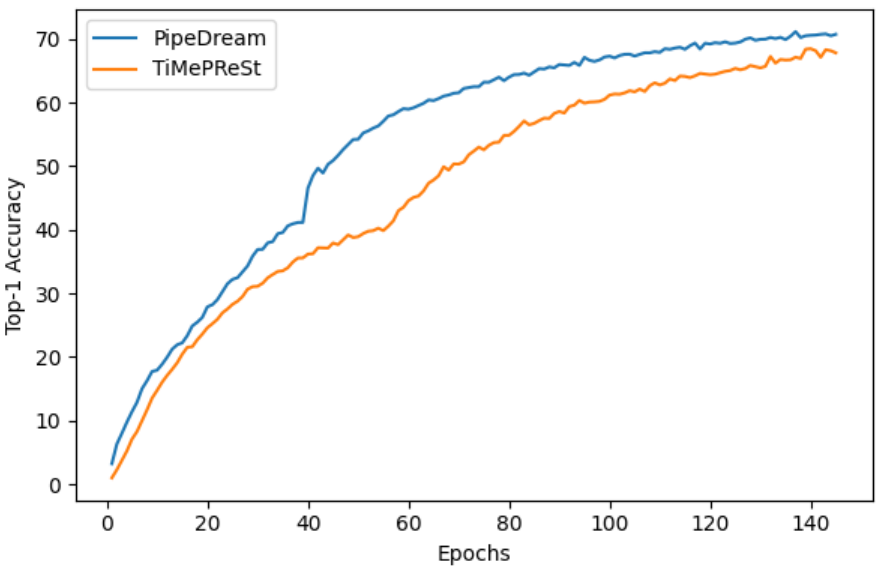}
    \caption{Top-1 accuracy to epoch}
    \label{fig:Top-1 accuracy to time_TinyImageNet200}
  \end{subfigure}
  \begin{subfigure}{.31\textwidth}
  \centering
    \includegraphics[width=1.0\linewidth]{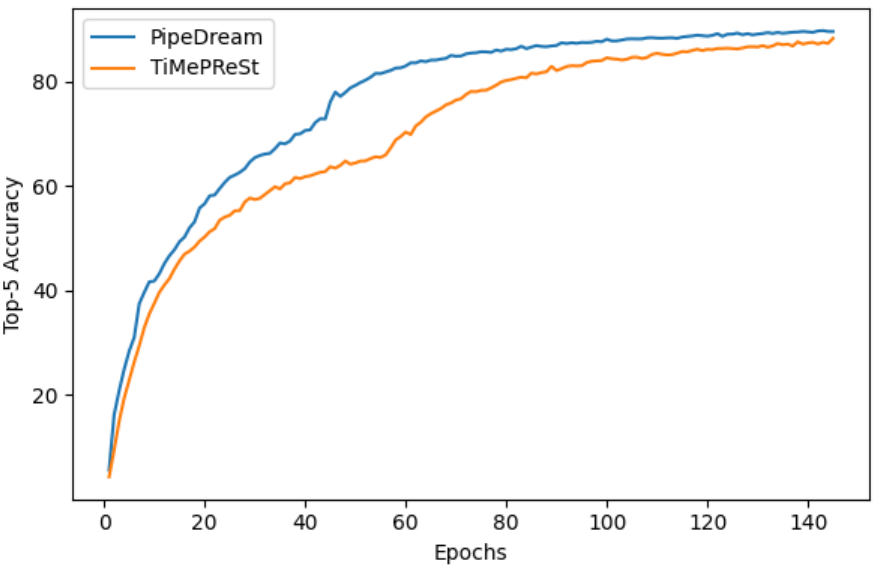}
    \caption{Top-5 accuracy to epoch}
    \label{fig:Top-5 accuracy to time_TinyImageNet200}
  \end{subfigure}
  \begin{subfigure}{.33\textwidth}
  \centering
    \includegraphics[width=0.99\linewidth]{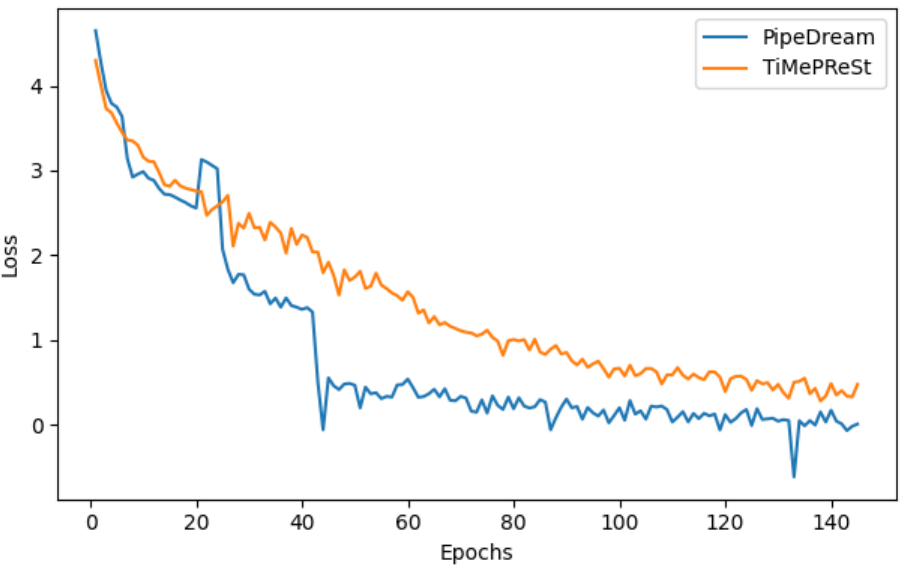}
    \caption{Loss comparison}
    \label{fig:Loss comparison_epoch_TinyImageNet200}
  \end{subfigure}
  \caption{\textbf{Statistical efficiency comparison of TiMePReSt and PipeDream (VGG-16 on Tiny-ImageNet-200)}}
  \label{fig: Statistical Efficiency Comparison of Proposed Method and PipeDream (VGG16 on Tiny-ImageNet-200)}
\end{figure}

\paragraph{\textit{Hardware efficiency, statistical efficiency and training time: }\normalsize\normalfont{Figures \ref{fig:Top-1 accuracy to time_cifar100} and \ref{fig:Top-5 accuracy to time_cifar100} clearly show that TiMePReSt can achieve a particular target accuracy in significantly less clock-time but with more iterations (epochs) than PipeDream. Since TiMePReSt achieves better hardware efficiency (time needed per epoch)\cite{narayanan2019pipedream} than PipeDream (Figure \ref{fig:Time per epoch}), thus, overall training speed-up is achieved in a significant scale although statistical efficiency (number of epochs needed to achieve a particular accuracy)\cite{narayanan2019pipedream} is compromised. In TiMePReSt, gradient calculation on latest updated version of weights rather than the weights used in forward pass, which have become stale over time, eliminates the need of horizontal weight stashing. Thus, GPU memory overhead is reduced, although statistical efficiency is compromised. However, to prevent the DNN from taking more clock-time to reach a target accuracy, the proposed intra-batch parallelism scheme of forward pass and avoiding backward passes for individual micro-batches helps improving hardware efficiency, thereby, overall training time.}}

\begin{figure}[h]
  \centering
   \includegraphics[width=0.8\linewidth]{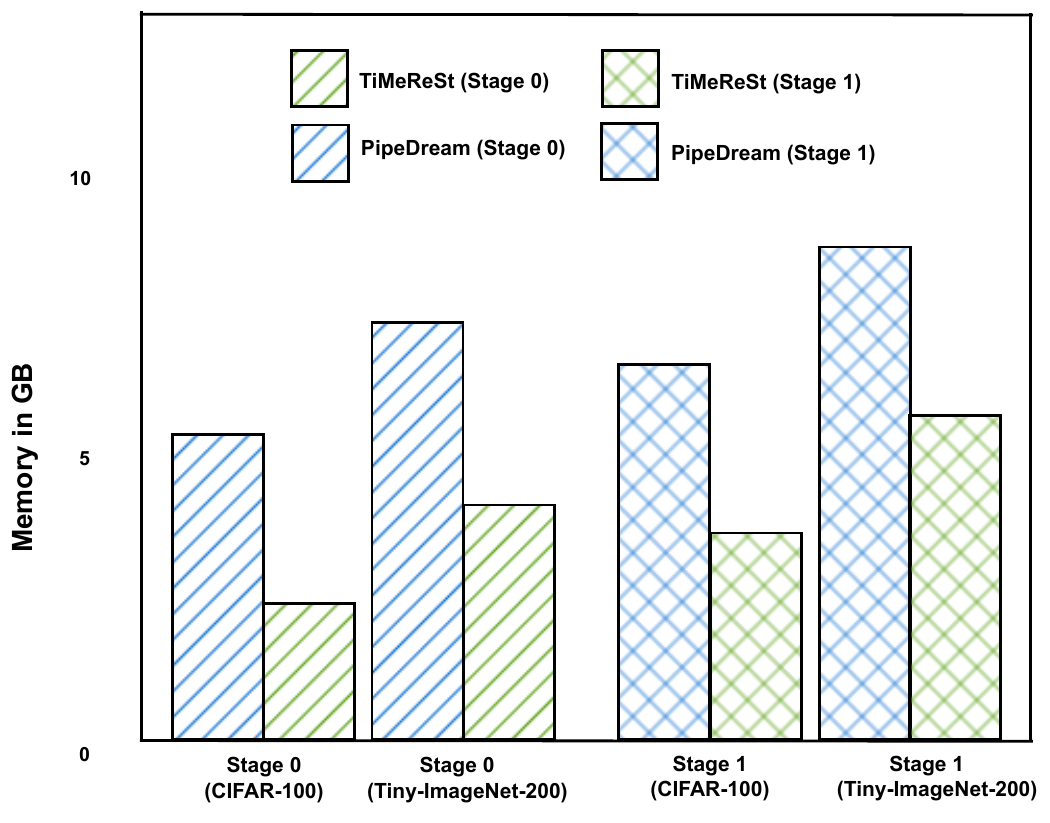}
  \caption{\textbf{Memory footprint per stage using two GPUs for training VGG-16 on CIFAR-100 and Tiny-ImageNet-200 image classification dataset, for TiMePReSt and PipeDream.}}
  \label{fig:Memory consumption comparison}
\end{figure}
\paragraph{\textit{Memory Overhead: }\normalsize\normalfont{Memory footprint is directly proportional to the number of weights and activations to be stashed \cite{narayanan2019pipedream}. Our experiment supports the fact by reducing memory footprint with the elimination of horizontal weight stashing. Figure \ref{fig:Memory consumption comparison} shows that TiMePReSt consumes nearly 50 percent and 40 percent less GPU memory in stage 0 and stage 1, respectively, than PipeDream for CIFAR-100, as well as for Tiny-ImageNet-200.}}
\section{Conclusion}
\label{Conclusion}
In this article, we have developed a methodology, called TiMePReSt, for pipeline-parallel DNN training, which helps reduce the GPU memory overhead and achieve zero degree of staleness of weight parameters by eliminating the need for horizontal weight stashing. GPU memory overhead acts as a bottleneck in training large DNNs on commodity hardware rather than costly high-end machines. TiMePReSt also reduces DNN training time significantly in pipeline-based frameworks by introducing an extra level of parallelization during forward passes, without increasing the frequency of backward passes, since backward pass is more computationally intensive than forward passes. Compared to PipeDream, TiMePReSt achieves a target accuracy 
much faster and reduces GPU memory consumption. TiMePReSt also achieves better hardware efficiency than PipeDream, resulting in higher throughput.
\\\\The key contributions of this paper are in the parallel training of an already partitioned DNN. However, this paper does not address the problem of efficiently partitioning any DNN model into phases across homogeneous or heterogeneous GPUs. A few previous attempts \cite{narayanan2019pipedream}\cite{fan2021dapple, zhao2022bapipe, park2020hetpipe} have worked on it. In the experiments, the DNNs have been partitioned roughly to ensure nearly optimal utilization of all the GPUs. As mentioned earlier, the cluster size is two in this work. Larger cluster can be formed, although there is no straight-forward relationship between training time and number of machines. More precisely, training time cannot be said to be inversely proportional to the size of a cluster. Network bandwidth between each pair of machines is one of the factors, which affects inter-machine communication time, resulting in overall training time. However, identification of the above mentioned relationship forms a future scope in this regard. The paper identifies a mathematical relationship between the number of micro-batches and worker machines in TiMePReSt and devises a mathematical expression for version difference, allowing for computation of different combinations without diagram preparation. 
\bibliography{Bibliography}
\end{document}